\documentclass[prb,aps,showpacs,amsmath,amssymb,amsfonts,preprint]{revtex4-1}
\usepackage{amsfonts}
\usepackage{amsmath}
\usepackage{amssymb}
\usepackage[dvips]{graphicx}%
\usepackage{graphicx}
\usepackage{float}
\usepackage{epsfig}
\usepackage{subfigure}

\begin{document}

\title{Elastic and magnetic properties of cubic Fe$_{4}$C from first-principles }%

\author{Gul Rahman}%
\email{gulrahman@qau.edu.pk}
\author{Haseen Ullah Jan}%
\affiliation{Department of Physics,
Quaid-i-Azam University, Islamabad 45320, Pakistan}

\begin{abstract}
First-principles based on density functional theory is used to study the phase stability, elastic, magnetic, and electronic properties of cubic (c)-Fe$_4$C. Our results show that c-Fe$_{4}$C has a ferromagnetic (FM) ground state structure compared with antiferromagnetic (AFM) and nonmagnetic (NM)states. To study the phase stability of c-Fe$_4$C, BCC  Fe$_4$C,  FCC  Fe$_4$C, and BCC  Fe$_{16}$C, where C is considered at tetrahedral and octahedral interstitial sites, are also considered. Although, the formation energy of c-Fe$_4$C is smaller than  BCC Fe$_4$C, but the shear moduli of c-Fe$_4$C is negative in the FM and AFM states indicating that c-Fe$_4$C is dynamically not stable in the magnetic (FM/AFM) states. However, NM state has positive shear moduli which illustrates that instability in c-Fe$_4$C is due to magnetism and can lead to soft phonon modes.
The calculated formation energy also shows that c-Fe$_4$C has higher formation energy compared with the FCC Fe$_4$C indicating no possibility of c-Fe$_4$C in low carbon steels at low temperature. The magnetic moment of Fe in  c-Fe$_4$C is also sensitive to lattice deformation. The electronic structure reveals the itinerant nature of electrons responsible for metallic behavior of c-Fe$_4$C.

 \end{abstract}

\maketitle

\section{Introduction}

The mechanical properties of steels can be improved by various types of solute atom,
which helps to pin the dislocations.
The base ingredient of steel is Fe, depending on the temperature,
adopts different crystal structures. 
For example,
at $912$\,K, $\alpha$-Fe, \textit{i.e.}, body centred cubic (BCC) transforms to $\gamma$-Fe face centred cubic (FCC), which further
transforms to $\delta$-Fe at $1700$\,K.
Light elements, like H, C, and N are considered to be playing a significant role in the phase stability and mechanical strength of Fe. Compared to the host Fe atom, radii of light atoms are smaller so
they do not substitute them and prefer to sit at the the interstitials sites.\cite{oct}
However, recent reports show that 
at the surfaces 
of both $\alpha$-Fe and $\gamma$-Fe,
C at the substitutional sites significantly reduces the surface energy~\cite{Harry, Korea}
so it prefers substitutional over the interstitials.

Iron carbides are of crucial importance in the design of the mechanical properties of steels and other iron alloys.\cite{a,b,c,d,f}
Carbon interstitials
are also present in various iron carbide phases (Fe$_2$C, Fe$_3$C, Fe$_7$C$_3$, Fe$_{16}$C) as precipitates in a ferrite matrix  or in pearlite-type or banite-type microstructures~\cite{a,b,c,d}.
Experimental investigations~\cite{Entin1973,Kaplow1983}
of martensites have indicated that the octahedral interstitial sites are preferred locations for carbon atoms in the BCC iron structure. The tetragonal distortion of the martensitic lattice is generally regarded as direct evidence of an unequal distribution of carbon atoms over the three available octahedral interstitial sublattices. When the carbon atoms occupy a single sub-lattice within an otherwise BCC iron lattice, the microscopic lattice symmetry is reduced to tetragonal, and a tetragonal distortion of the BCC iron lattice results. Zener~\cite{Zener1946} treated the thermodynamics of such carbon ordering, and the tetragonal martensite  is accordingly referred to as ``Zener-ordered.''
It was shown~\cite{Izotov1968,Choo1973,Ino1982,Nagakura1979} that a secondary ordering of carbon interstitials may occur within the Zener sublattice during martensite aging, prior to the detection of  $\epsilon$- or $\epsilon^{\prime}$-carbide. These results have been interpreted in terms of the formation of the Fe$_{4}$C phase with a considerable deviation from stoichiometric composition.\cite{Ino1982}

\begin{figure}
\centering
\subfigure[]{\label{}\includegraphics[width=0.251361253\textwidth]{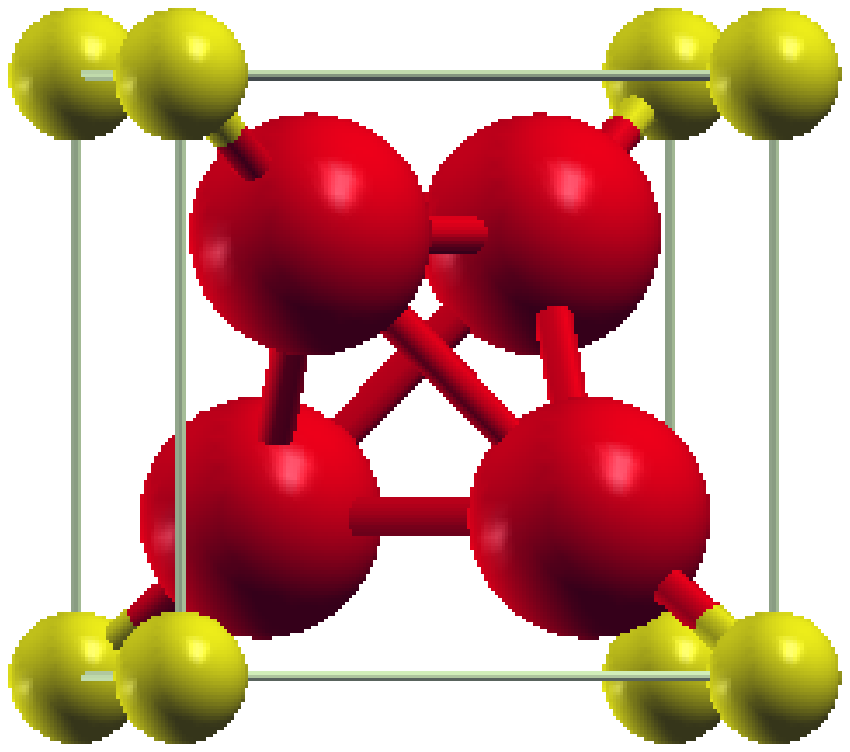}}                
\subfigure[]{\label{}\includegraphics[width=0.2513612653\textwidth]{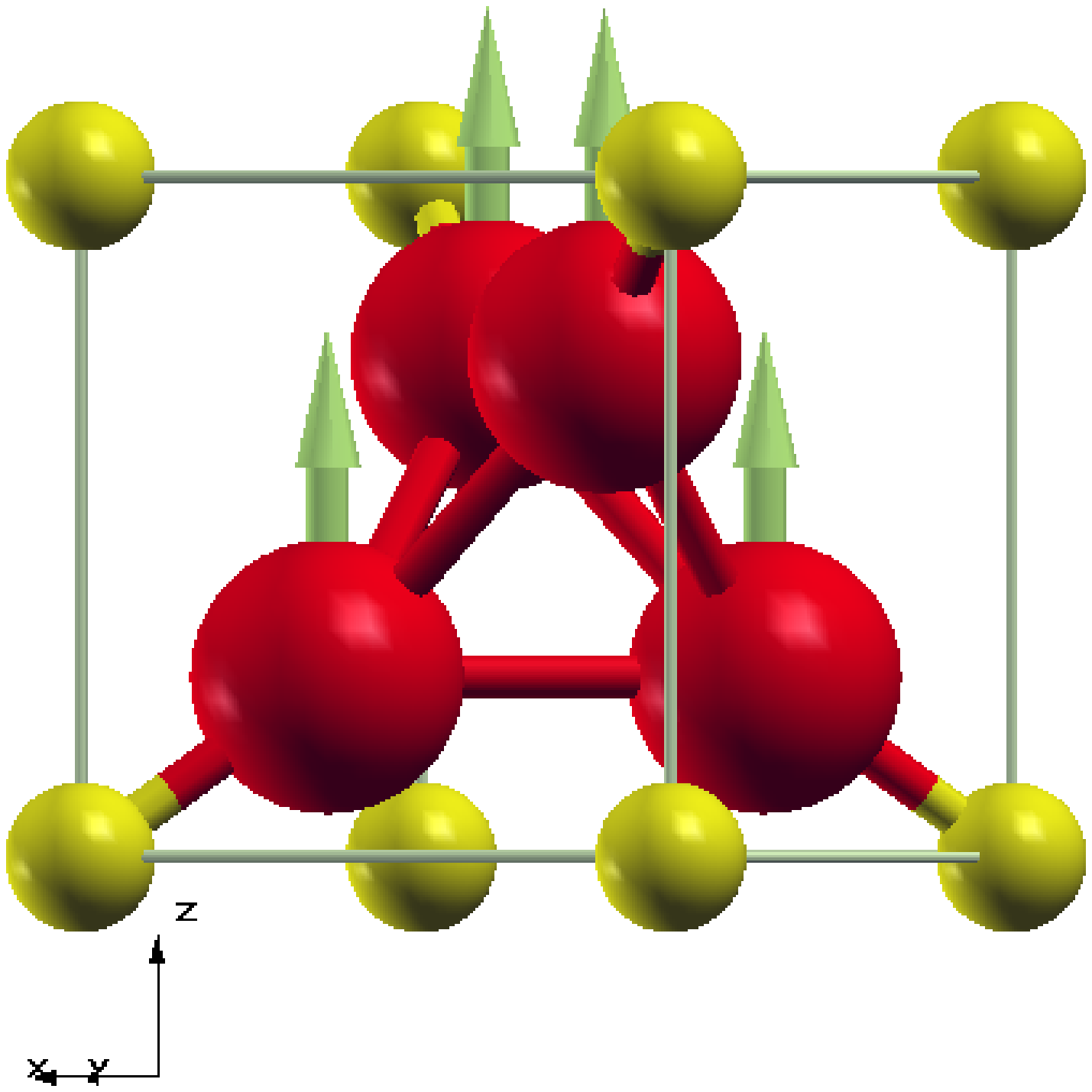}}
\subfigure[]{\label{}\includegraphics[width=0.2451361243\textwidth]{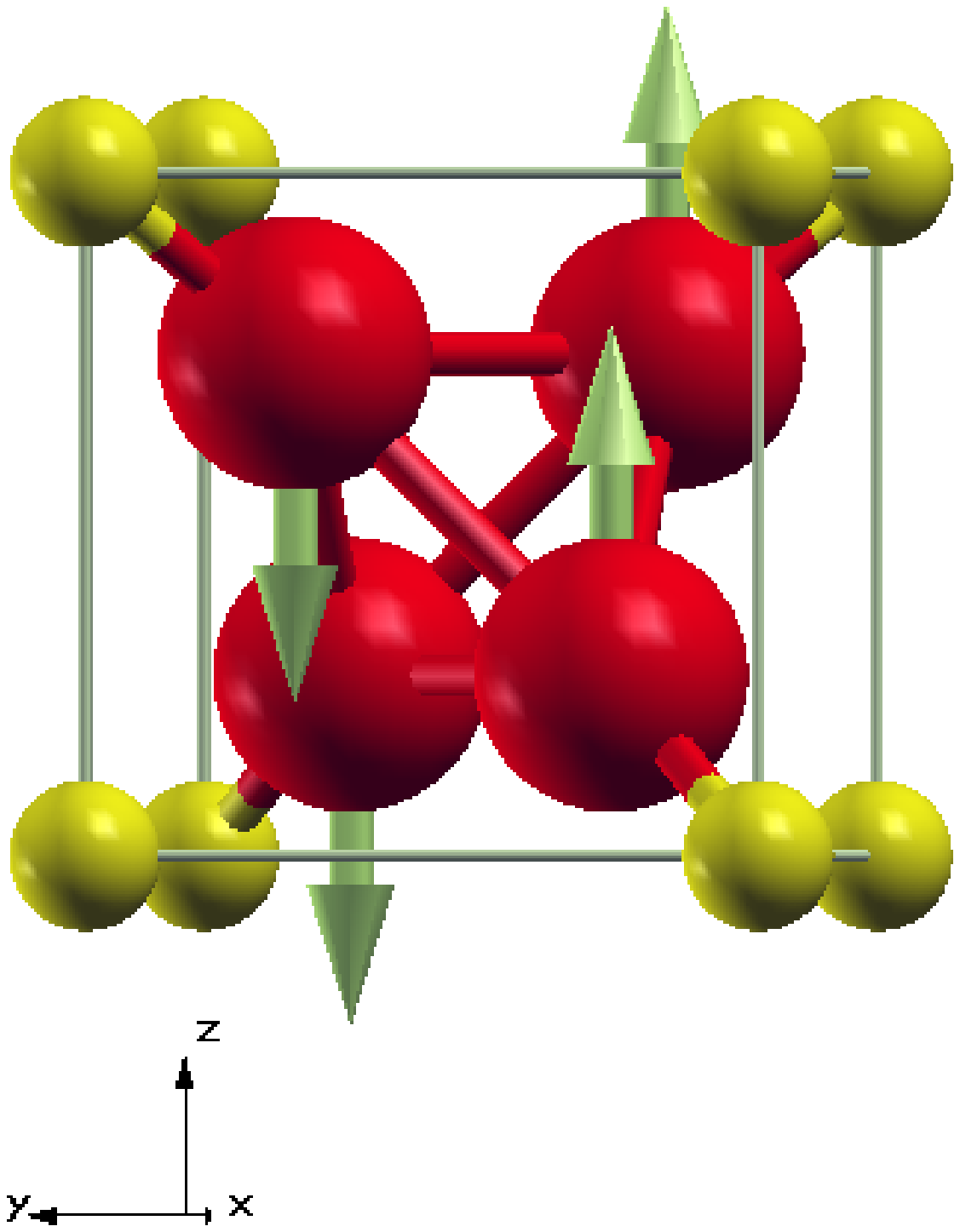}}
\caption{(Colour online) Crystal structures of c-Fe$_{4}$C in the NM (a), FM (b), and AFM (c) states. The up and down arrows show the direction of spin magnetic moments. Red (yellow) balls show Fe (C) atoms.}
\label{nafm}
\end{figure}

In the past much attention has been focused to interstitial C in BCC Fe, \textit{i.e.}, Fe$_4$C and FCC Fe, \textit{i.e.}, Fe$_4$C.\cite{CPRB} 
There is an other possible phase of Fe-C system---cubic Fe$_4$C denoted as c-Fe$_4$C (see Fig.\ref{nafm}). 
In this phase, the C atoms are not located at interstitial sites, but at the regular  lattice sites.
There are four Fe atoms and one C atom per unit cell and 
has a lattice constant of $3.89$\,\AA .\cite{http://cst-www.nrl.navy.mil/lattice/struk/Fe4C.html}
  \,\,This structure has a pearson symbol ``cP5"
and belongs to the space group ``P43m".
The fractional atomic positions of Fe and C atoms are given in Table\,\ref{AtomPosition}.
Due to its very different crystal structure than those of other iron carbides, 
it is very important to investigate its physical properties from first-principles and to see any possibility of formation in steels, because there is no theoretical work has been done on c-Fe$_{4}$C.
We study the phase stability, elastic, magnetic, and electronic properties of c-Fe$_{4}$C. Three different magnetic states, \textit{i.e.}, non-magnetic (NM), ferromagnetic (FM), and anti-ferromagnetic (AFM) states (see  Fig.\ref{nafm})  of the c-Fe$_4$C are considered.   
To compare the structural properties of c-Fe$_4$C, we also considered BCC Fe$_4$C, Fe$_{16}$C and FCC
Fe$_4$C. These results will be helpful for understanding the relative stability of iron and iron carbides in steels.

\section{ Calculation Methods}

\subsection{Computational Details}

  
 All calculations were carried out using the {\sc Siesta}~\cite{Ordejon1996} $\textit{ab initio}$ density functional theory\,\cite{KSH} code.
We used the spin-polarized Perdew-Burke-Ernzerhof (PBE) version\,\cite{PBE} of the Generalised Gradient Approximation 
(GGA) to the exchange and correlation functional.
Using Troullier-Martins parametrization~\cite{TMP} and one projector for each \textit{l} state, core electrons are described by norm-conserving pseudopotentials.
We used double-$\zeta$ polarized (DZP)
basis sets optimised for solid Fe and C.
Different meshes were used for the real space discretization. Finally, a mesh cut off of 400\,Ry for real space discretisation and an $m\,\times\,m\times\,m$ ($m = 8$) Monkhorst-Pack grid~\cite{Ordejon1996} for Brillouin zone (BZ) sampling were used.
The latter corresponds to $\textit{k}$ points in the irreducible part of the BZ.
Both lattice parameters and coordinates of atoms were optimized.

\subsection{Method for formation energies }
 The formation energy ($\Delta$E) of an iron carbide Fe$_{n}$C$_{m}$ is described as,
\begin{equation}\label{FE}
\Delta E = E(\rm{Fe}_\mathrm{n}\rm{C}_\mathrm{m}) - [n E(\rm {Fe}) + m E(\rm {C})]
\end{equation}
In the above equation $E(\rm{Fe}_\mathrm{n}\rm{C}_\mathrm{m})$ is the total energy of FeC systems. Similarly $E(\rm {Fe})$ and $E(\rm {C})$ represent the total energy of elemental solids, where $n$ and $m$ are number of  Fe and C atoms, respectively.  
We considered both the $\alpha$ and $\gamma$ phases of Fe, and diamond C
for the evaluation of the formation energy.
The relative stability of c-Fe$_\mathrm{4}$C with respect to various iron carbides 
is determined by the formation energy per atom, given by

\begin{equation}\label{FEPA}
\Delta E_{f} = \Delta E/\rm{(n + m)}
\end{equation}
Here, negative values of $\Delta\textit{E}_\mathrm{\textit{f}}$ means that Fe$_\mathrm{n}\rm{C}_\mathrm{m}$ is more stable than the parent solids.

\subsection{Method for Elastic Properties}
Elastic properties can better be understood by calculating elastic constants $C_\mathrm{ij}$. For this, volume conserved strain matrices~\cite{GUL} are applied to deform the lattice. A $\pm$0.03 amount of strain ( $\varepsilon$) is imposed on the cubic (c), BCC, and FCC phases of Fe$_{4}$C, where
different magnetic states, \textit{i.e.}, NM, FM and AFM are considered.
The total energies were calculated as a function of lattice volume and fitted to the Birch-Murnaghan equation of state (EOS),\cite{EOS} to find the optimized lattice volumes and the bulk moduli. 
 Cubic systems are well described by three independent elastic constants, \textit{i.e.,} $C_\mathrm{11}$, $C_\mathrm{12}$, and $C_\mathrm{44}$, where these elastic constants represents deformations along different diagonals in a unit cube of the lattice.
 The three elastic constants are determined by calculating the total energy as a function of the bulk volumetric shears.\cite{GUL}

\section{RESULTS and Discussions}
  
\subsection{Formation energies and phase stabilities}   

Formation energy is a measure of the phase stability of a system. 
Before calculating the formation enthalpy, we first optimized the lattice parameters for the $\alpha$ and $\gamma$ phases of Fe,
 diamond C, and Fe$_{4}$C in different phases, \textit{i.e.}, cubic , BCC, and FCC in different magnetic states,
\textit{i.e.}, NM, FM, and AFM.

\begin{figure}
\begin{center}
\includegraphics[width=0.5\textwidth]{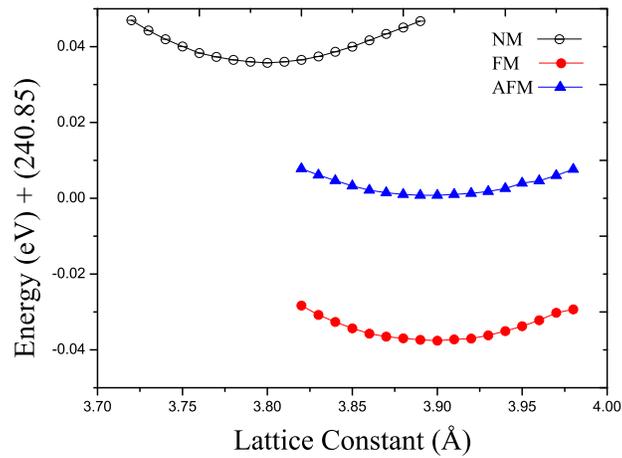}
\caption{(Colour online) Calculated total energy vs lattice constant in different magnetic states of c-Fe$_4$C. The empty circles, filled circles, and filled triangles represent the NM, FM, and AFM states of c-Fe$_4$C, respectively.}
\label{phase}
\end{center}
\end{figure}

Our optimized lattice parameters are presented in Table \ref{t1}. 
We see that our calculated lattice parameters of Fe and C are in good agreement with the previous theoretical and experimental work.\cite{Kittle,FeCPRB,CPRB}
The c-Fe$_{4}$C expands by about $2.63\%$ in both the magnetic phases as compared to the non-magnetic phase. 
This magneto-volume effect has significant influence on the elastic and magnetic properties 
as discussed shortly. The representative binding energy curves of NM, FM, and AFM states for the c-Fe$_{4}$C are shown in Fig.\ref{phase}. It is clear from this figure that FM has the lowest energy with lattice constant 3.89\,\AA. Therefore our DFT calculations show that FM is the magnetic ground state.
The calculated lattice constants for the BCC and FCC Fe$_{4}$C are 2.93\,\AA\, and 3.82\,\AA\,, respectively, in the FM ground states. Previous results for the FCC Fe$_{4}$C are 3.76\,\AA\,\cite{CPRB} and 3.87\,\AA\,\cite{FEC} in theory and experiment,the present calculated values are slightly smaller than those obtained at room temperature, as might be expected from thermal expansion, but it is encouraging that the trend of the lattice parameter as compared with elemental solids (Fe, C) are correctly reproduced (see Table\,\ref{t1}).

Manipulating the optimized lattice parameters, formation energy is calculated using Eq.\eqref{FE} and the results are  tabulated in Table\,\ref{t2}. Formation enthalpy is nothing more than the total energy of a system at zero pressure and zero kelvin at the corresponding equilibrium lattice constant. The formation enthalpy is measured relative to BCC Fe, FCC Fe, and diamond C for all the Fe-C systems.
For the purpose of phase stability, we also considered BCC Fe$_4$C and Fe$_{16}$C, where C atom is considered at octahedral and tetrahedral positions. BCC Fe$_{4}$C and Fe$_{16}$C were model as 1$\times$\,1$\times$\,2 and 2\,$\times$\,2$\times$\,2 supercells, respectively, and the C atom was considered at octahedral and tetrahedral sites. Our calculated formation enthalpy for the Fe$_{16}$C at the octahedral and tetrahedral sites with respect to the BCC Fe lattice are 32.057 and 32.062\,kJ/mol, respectively (see Table\,\ref{t2}), which are in good agreement with the previous work, \textit{i.e.}, 20.19\,kJ/mol at the octahedral and 24.89\,kJ/mol at the tetrahedral site,\cite{HKDH.2013} also $\Delta E_f$ for the BCC Fe$_4$C is 116.363kJ/mol at the octahedral,  and 100.830\,kJ/mol
at the tetrahedral lattice sites.
The formation enthalpy at both the octahedral and tetrahedral Fe$_4$C is larger than Fe$_{16}$C due high carbon concentration. 
Table\,\ref{t2} clearly shows that FCC Fe$_{4}$C has the lowest formation energy, \textit{i.e.}, -3.751\,kJ/mol with respect to the FCC Fe lattice and highest one, \textit{i.e.}, 89.899\,kJ/mol with respect to the BCC Fe lattice. It shows that FCC Fe$_{4}$C with respect to the FCC Fe lattice is the most stable phase among these Fe-C systems. This is also in agreement with the experimental observations.\cite{FeCPRB} 
In all these interstitial carbides, we see that Fe$_n$C$_m$ has the lowest formation enthalpy with respect to  BCC Fe lattice.
It is noticeable that c-Fe$_{4}$C has lower formation energy than BCC  Fe$_{4}$C with respect to both, \textit{i.e.}, BCC and FCC Fe lattices, however this is dynamically not a stable structure which will be discussed below. This indicate, that it will not be easy for C to form such c-Fe$_{4}$C structure in low carbon steels.

\subsection{Elastic properties}
In order to obtain the elastic properties, we followed our previous approach~\cite{GUL} and $\pm$0.03 amount of strains was imposed, using the shear and elastic volume conserved deformation matrices. The calculated bulk moduli and elastic constants are given in Table\,\ref{CB}. The calculated elastic constants of BCC Fe and diamond C are close to the experimental values~\cite{Kittle} and confirm that the number of basis functions and the number of $k$-points used in these calculations are sufficient to reproduce the experimental data for Fe and C. It is noticeable that BCC Fe and  diamond C have positive shear modulus, whereas FCC Fe has negative one which indicates that FCC Fe is dynamically unstable and BCC Fe is the most stable structure of Fe. Also note that the FCC Fe$_{4}$C has positive elastic constants, which further demonstrate the phase stability of C interstitials in FCC Fe. This is also very encouraging that C free FCC Fe is not stable, however it becomes stable when C is placed at the octahedral interstitial site. This phase stability is accompanied by the volume expansion (see Table\,\ref{dlattice}), and reduction of magnetization which will be discussed below. We may also infer from these calculations that the FCC Fe either can be stabilized in the form of a thin film on a substrate, \textit{e.g.}, Cu or by inserting C at the interstitial sites.\cite{Growth of stabilized g-Fe ﬁlms and their magnetic properties} Moreover, the formation energy of FCC Fe$_{4}$C with respect to BCC Fe is found to be 89.899 kj/mol which further illustrates the formation of FCC Fe$_4$C over BCC Fe$_4$C.

\begin{figure}
\begin{center}
\includegraphics[width=0.5\textwidth]{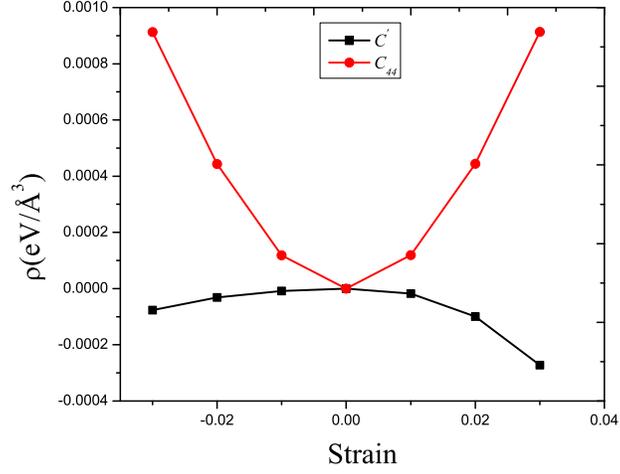}
\caption{(Colour online) Calculated energy density ($\rho$) vs strain for the shear modulus $\acute{C}$ (black filled squares) and elastic constant C$_{44}$ (red filled circles) of c-Fe$_4$C in the FM state.}
\label{Fe4CCpC44}
\end{center}
\end{figure}

To discuss the elastic properties of c-Fe$_{4}$C, we calculate elastic constants $C_{ij}$ in the NM, FM, and AFM states.
The representative curves of $C_{ij}$ in the FM state are shown in Fig.\ref{Fe4CCpC44}. In the magnetic (FM/AFM) states c-Fe$_{4}$C has negative curvature, but not in the NM case (see Table\,\ref{CB}). This shows that c-Fe$_{4}$C is dynamically not stable due to magnetism. In other words, c-Fe$_{4}$C may be stable at a very high temperature due to thermal expansion and the high temperature will destroy the magnetic state of c-Fe$_{4}$C as well.
To further explore the elastic properties of c-Fe$_{4}$C, we also calculated $C_{ij}$ at different lattice constants, \textit{i.e.}, 3.89\AA\,---\,3.85\AA\, (volumes) in the FM state and the results are summarized in Table\,\ref{dlattice}. It is clear to see that the absolute values of $C_{12}$ and $C_{44}$ increase, where as and $C^{\prime}$ and $C_{11}$ decrease with decreasing volume. Note that $C^{\prime}$ remains negative in the whole volume (lattice) range. This clearly indicates that the phase instability is coming from the magnetic interactions of C atom with the nearest Fe atoms.
This shows that magnetization mainly helps to stabilize and destabilize ferrous materials.\cite{GUL, OIGorbatov2013}

\subsection{ Magnetism}

\begin{figure}
\begin{center}
\includegraphics[width=0.5\textwidth]{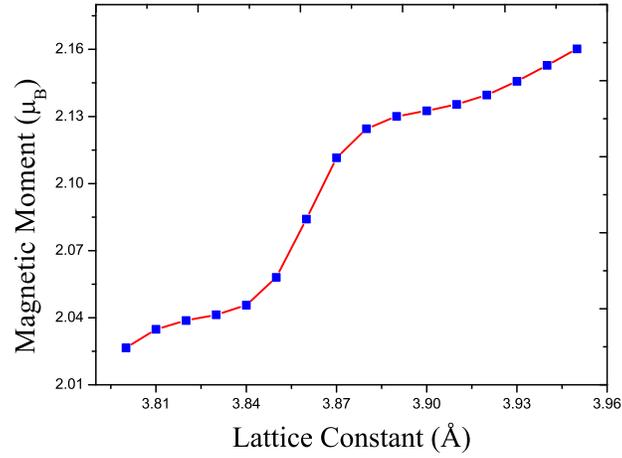}
\caption{(Colour online) The calculated total magnetic moment ($\mu_B$) per Fe atom vs lattice constant (\AA) of c-Fe$_4$C.}
\label{scmm}
\end{center}
\end{figure}

We calculated the magnetic moment per Fe atom for the BCC and FCC Fe, Fe$_{4}$C and the results are summarized in Table\,\ref{t5}. Our calculated values agree with the the available experimental work. The magnetic moments of pure BCC and FCC Fe are 2.24\,$\mu_B$ and 3.59\,$\mu_B$, respectively and are in good agreement with experiments 2.22\,$\mu_B$ (BCC Fe)\cite{Kittle} and theory 3.36\,$\mu_B$ (FCC Fe).\cite{FEC} In c-Fe$_{4}$C each Fe atom contributes approximately equal magnetic moment (2.11\,$\mu_B$) due to the same lattice symmetry. On the other hand BCC Fe\,$_{4}$C has 1.77\,$\mu_B$ per Fe atom, where FCC Fe$_{4}$C has 2.04\,$\mu_B$ per Fe atom. The previous theoretically calculated magnetic moment for the FCC Fe\,$_{4}$C is 2.98\,$\mu_B$ per Fe atom.\cite{FEC}  
We see that the magnetic moments are slightly decreased in Fe$_{n}$C$_{m}$.
We also considered the effect of lattice volume on the magnetic moments of c-Fe$_{4}$C, which are presented in Fig.\ref{scmm}. This figure shows that magnetic moment increases with lattice volume  as found in the other Fe-based systems.\cite{DSantos.MMM.1997} This increment is mainly due to weak hybridization effect at higher volume. We can see an abrupt change in the magnetic moments around $3.85$---$3.90$\,\AA.
This abrupt change in magnetic moment indicates that c-Fe$_{4}$C is very sensitive to lattice deformation which can be lead to the phase stability of the system. This abrupt change may be due to spin-flip effect.\cite{DSantos.MMM.1997} 

\begin{figure}
\begin{center}
\includegraphics[width=0.5\textwidth]{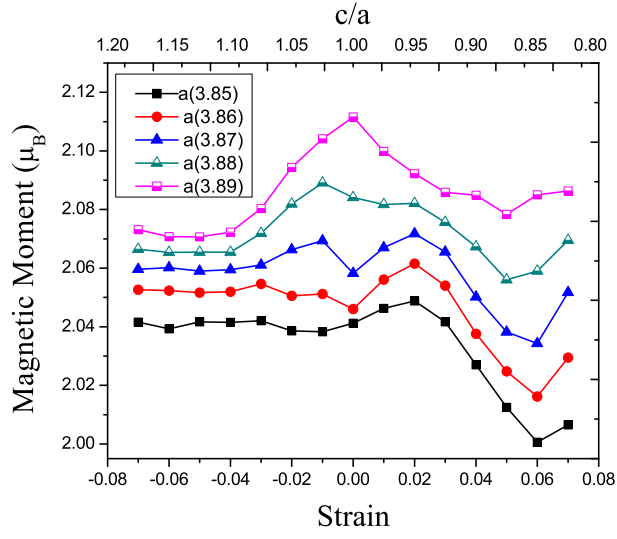}
\caption{(Colour online) Strain($c/a$) vs calculated magnetic moment ($\mu_B$) per Fe atom of c-Fe$_4$C.
Filled squares (black), filled circles (red), filled triangles (blue), half filled triangles (misson), and half filled squares (magenta ) represent lattice constants from 3.85---3.89\AA\ of c-Fe$_4$C.}
\label{cpmm}
\end{center}
\end{figure}

In de-localized electron systems, magnetism originates from the difference in the spin up and spin down bands' populations. Also deformation affects de-localized systems drastically due to strong interaction of magnetic moments and it would be very necessary to study the effects of deformation.
As shown above that c-Fe$_{4}$C is dynamically not stable in the FM/AFM state. To further investigate the effect of strain on magnetism, we applied $\pm 0.07$ volume conserved strains for different lattice constants, which was varied from $3.89$\,\AA\,\,to $3.85$\,\AA. The results are summarized in Fig.\ref{cpmm}. The magnetic moment decreases with decreasing lattice volumes consistent with Fig.\,\ref{scmm}, \textit{i.e.}, magnetic moment is maximum at the optimized lattice constant (3.89\,\AA\,) and decreases up to (3.85\,\AA\,). At each lattice constant, the magnetic moment of the un-strained system is maximum and decreases as the strain increases. The magnetism of c-Fe$_{4}$C is very sensitive around $c/a=1.0$, and at large elongation the magnetic moments of c-Fe$_{4}$C are saturated, whereas at large compressions the magnetic moment decreases. Also, the maximum in the magnetic moment shifts to smaller value of c/a at smaller volume of c-Fe$_{4}$C. 
Such magnetic behaviour clearly indicates that  c-Fe$_{4}$C is dynamically unstable due to magnetism.

\subsection{Electronic structure}
\begin{figure}
\begin{center}
\includegraphics[width=0.31\textwidth]{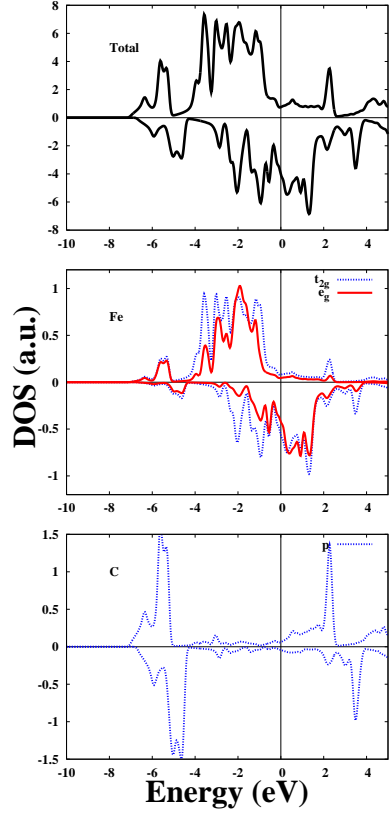}
\caption{(Colour online) Calculated total density of states (DOS) for c-Fe$_{4}$C in FM state, and partial density of states (PDOS) of Fe and C atoms. Solid (red) and dashed (blue) lines represent e$_{g}$ and t$_{2g}$ states of Fe. Bottom panel of blue dotted lines represent C $p$ states. Top panel solid line represents total DOS. The positive (negative) DOS shows majority (minority) spin states. The Fermi level (E$_\mathrm{F}$) is set to zero.}
\label{scdos}
\end{center}
\end{figure}

\begin{figure}
\begin{center}
\includegraphics[width=0.4\textwidth]{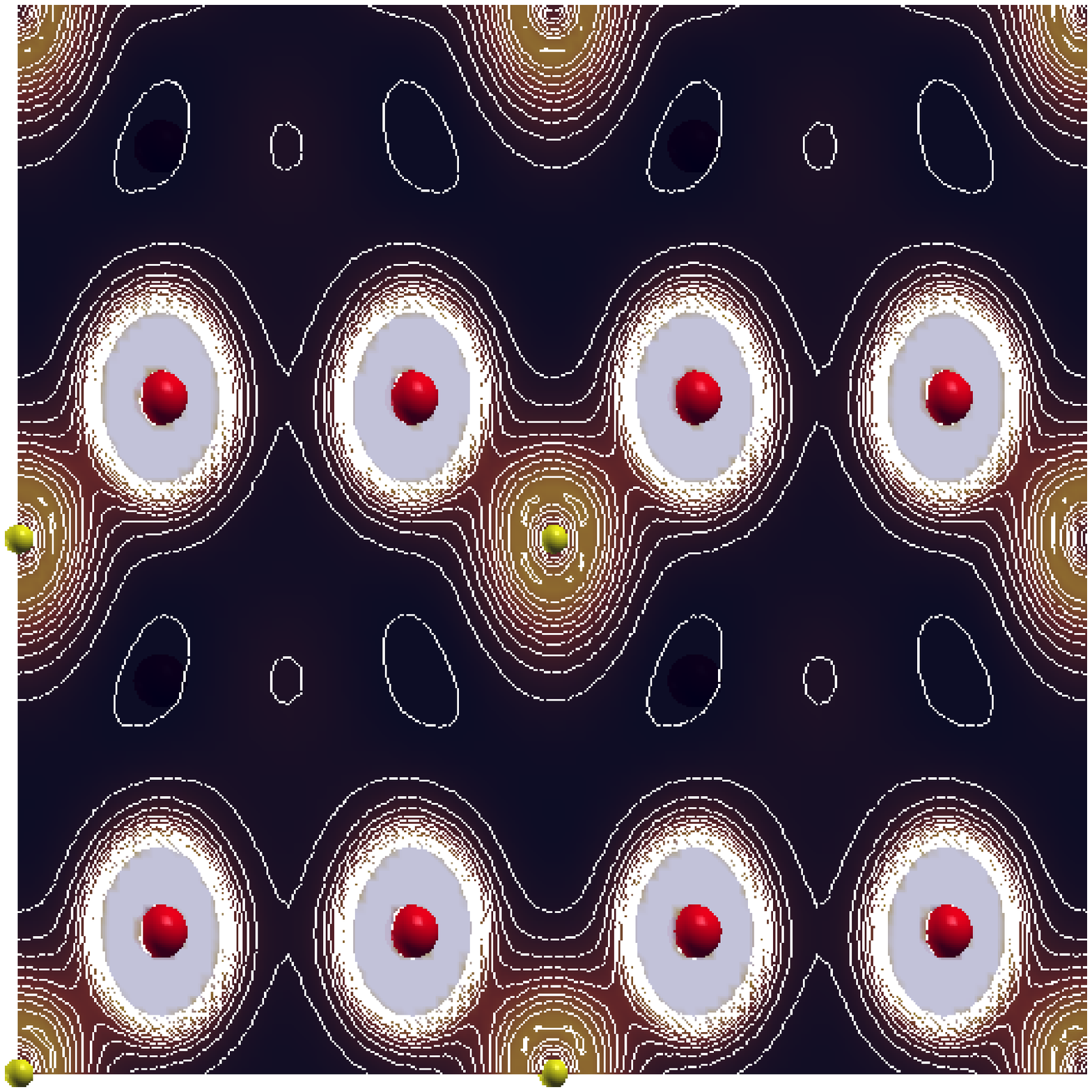}\hspace{0.5 cm}
\includegraphics[width=0.2\textwidth]{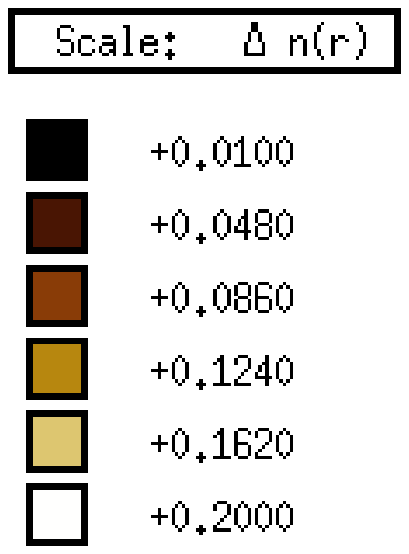}
\caption{(Colour online) Charge density contour along the (110) plane of the c-Fe$_4$C. Red and yellow balls represent Fe and C atoms, respectively. The cut-off plane is passing through the second neighbour atoms. Darker and lighter regions corresponds to low and high
charge density. The scale is also shown.}
\label{sccharg}
\end{center}
\end{figure}

Elastic and magnetic properties and phase stability and instability ultimately have an electronic origin.
Fig.\ref{scdos} represents the total and atomic projected density of states (PDOS) in the FM state of c-Fe$_4$C. The total DOS of c-Fe$_4$C has low lying sharp peaks coming from the hybridization of C and Fe atoms. The DOS of c-Fe$_{4}$C shows that the $t_{2g}$ and $e_{g}$ states of Fe-$d$ are almost degenerate which is quite different from BCC/FCC Fe and BCC/FCC Fe$_{4}$C.  Near the Fermi energy, C and Fe atoms also hybridises and the DOS shows metallic behavior. The metallicity of c-Fe$_4$C consistent with the pure $\alpha$ and $\gamma$-Fe. The strong bonding nature of Fe-C in c-Fe$_4$C is also visible in charge density plot which is shown in Fig.\ref{sccharg}.  This strong bonding is responsible for the smaller magnetic moments of Fe in c-Fe$_{4}$C as compared with the 
BCC/FCC Fe.

\section{Summary} 
The elastic and magnetic properties of c-Fe$_4$C are studied using first-principles calculations. Different magnetic phases (NM, FM, and AFM) of c-Fe$_4$C were considered, and we found FM ground state for  c-Fe$_4$C. For the phase stability of  c-Fe$_4$C, BCC  Fe$_4$C,  FCC  Fe$_4$C, and BCC  Fe$_{16}$C, where C was located at tetrahedral and octahedral interstitial sites, were also considered. 
The formation energy of c-Fe$_4$C was smaller than  BCC Fe$_4$C, but the shear moduli of c-Fe$_4$C was negative in the FM and AFM states indicating that c-Fe$_4$C is dynamically not stable in the magnetic (FM/AFM) states.  In contrast, NM state showed positive shear moduli which illustrates that instability in c-Fe$_4$C is due to magnetism. The magnetic moment of c-Fe$_4$C increases with increasing lattice constant, similar to BCC Fe. The effect of lattice distortion on magnetic moments with different lattice constants were also studied and it was confirmed that magnetic moments decreases with strain. We concluded that c-Fe$_4$C can not form a stable phase in low carbon steels at low temperature. The electronic density of states and charge density revealed metallic behavior of electrons in c-Fe$_4$C. \\

\section*{ACKNOWLEDGMENT}
 G. R. acknowledges the cluster facilities of NCP, Pakistan.

\newpage

\begin{table}
\caption{Atomic fractional coordinates of c-Fe$_4$C structure.}
\begin{center}
\begin{tabular}{l c c c }
\hline\hline
Atom &$x$ &  $y$ & $z$  \\
\hline
Fe & 0.265 & 0.265 & 0.265 \\
Fe & 0.265 & -0.265 & -0.265 \\
Fe & -0.265 & 0.265 & -0.265 \\
Fe & -0.265 & -0.265 & 0.265 \\
C & 0.0 & 0.0 & 0.0 \\ 
\hline 
\hline
\end{tabular}
\label{AtomPosition}
\end{center}
\end{table}

\begin{table}
\caption{Calculated lattice constants for BCC and FCC Fe, diamond C, and Fe$_4$C
in different magnetic states and crystal structures. The lattice constants for NM, FM, and AFM
states are given in units of \AA.}
\begin{tabular}{l c r l}
\hline\hline
 System & This Work & & Experiment \\
BCC (Fe) & 2.86 & 2.86$^a$ & 2.83$^b$  \\
FCC (Fe) & 3.68 & 3.58$^b$ & 3.60$_b$ \\
Diamond (C) & 3.56 & 3.56$^a$ & 3.57$^c$ \\
c-Fe$_4$C(NM) & 3.79 & - & - \\
c-Fe$_4$C(FM) & 3.89 & - & -  \\
c-Fe$_4$C(AFM) & 3.89 & - & - -\\
BCC (Fe$_4$C)  &2.93 & - & - \\
FCC (Fe$_4$C) & 3.82 & 3.87$^d$ & 3.76$^c$   \\
\hline\hline
$a$ Ref.~\cite{Kittle}\\
$b$ Ref.~\cite{FeCPRB}\\
$c$ Ref.~\cite{CPRB}\\
$d$ Ref.~\cite{FEC}\\
\end{tabular}
\label{t1}
\end{table}

\begin{table}
\caption{ Calculated formation enthalpies per atoms ($\Delta E_f$) in units of (kJ/mol) with respect to BCC and FCC Fe for different phases of Fe$_4$C. TET and OCT stand for tetrahedral and octahedral interstitial sites.}
\begin{center}
\begin{tabular}{l c r }
\hline\hline
System &   BCC~$\Delta E_f$\, & FCC~$\Delta E_f$ \\
\hline
TET Fe$_{4}$C &100.830\,& 113.463  \\
TET Fe$_{16}$C &  32.062\,& 85.961 \\  
OCT Fe$_{4}$C & 116.363\,& 132.878 \\
OCT Fe$_{16}$C & 32.057\,& 85.940  \\ 
c-Fe$_{4}$C& 46.339\,& 45.348 \\ 
FCC Fe$_{4}$C & 89.899\,& -3.751 \\ 
\hline 
\hline
\end{tabular}
\label{t2}
\end{center}
\end{table}

\begin{table}
\caption{Calculated Elastic constants and bulk modulii in units of Giga Pascal (GPa) for different systems.}
\begin{center}
\begin{tabular}{cccccc}
\hline\hline
System Name\, & C$_{11}$\, & C$_{12}$\, & C$_{44}$\, & B&\,$\acute{C}$  \\
\hline
 BCC Fe\, &257.436\,& 137.216\,& 100.184\,&177.290&\,60.11 \\
 FCC Fe\, &67.762\,& 189.430\,& -46.610\,& 148.874 &\, -60.83 \\
 Diamond C\, & 1091.420\,& 182.612\,&571.482\,&485.548&\, 454.40  \\
 c-Fe$_{4}$C (FM)\, &-  & -  &80.300\,& 174.965&\, -4.07 \\
 c-Fe$_{4}$C (NM)\,&252.141\,&195.864\,&147.733\,&214.623&\, 28.13 \\
FCC-Fe$_{4}$C\, &400.204\,& 164.088\,&25.490\,& 242.793&\, 118.05\\
\hline\hline
\end{tabular}\\
\label{CB}
\end{center}
\end{table}

\begin{table}
\caption{The calculated elastic constants (GPa) and shear modulus $\acute{C}$ for c-Fe$_4$C, obtained at different lattice constants.}
\begin{center}
\begin{tabular}{l c c c r }
\hline\hline
\textit{a} (\AA)\, & C$_{11}$\, & C$_{12}$\, & C$_{44}$\, & $\acute{C}$ \\
\hline
 3.89\, & 146.15\,& 189.38 \,& 80.30 \,& -21.61\\
 3.88\, & 145.5 \,& 189.7  \,& 80.00\, & -22.09 \\
 3.87\, & 143.44 \,& 190.57\,& 81.03\, & -23.39 \\
 3.86\, & 141.72 \,& 191.6 \,& 83.85\, & -24.93 \\
 3.85\, & 140.62 \,& 192.15\,& 82.85\, & -25.76 \\
\hline\hline
\end{tabular}
\label{dlattice}
\end{center}
\end{table}

 \begin{table}
\caption{The DFT calculated total magnetic moment per Fe atom ($\mu_B$) for the BCC and FCC Fe, and different
crystal structures of Fe$_4$C in units of the Bohr magneton ($\mu_B$) are well compared with the available literature (theory and experiment). LS and HS in the subscripts stand for low spin and high spin, respectively.}
\begin{center}
\begin{tabular}{l c c c c }
\hline\hline
 System Name & This Work & Experiment & Theory \\
\hline
BCC (Fe) & 2.24 & 2.22$^a$ & 2.15$^b$ \\
FCC (Fe) & 3.59 & - & 0.126$^c$$_{LS}$, 3.359$^c$$_{HS}$\\
c-Fe$_4$C & 2.11 & - & -\\
BCC Fe$_4$C  & 1.77 & - & -\\
FCC Fe$_4$C & 2.04 & - & 2.98$^c$ \\
\hline\hline
$a$ Ref.~\cite{Kittle}\\
$b$ Ref.~\cite{FeCPRB}\\
$c$ Ref.~\cite{FEC}\\
\end{tabular}
\label{t5}
\end{center}
\end{table}

\clearpage

\end{document}